\documentclass[]{iopart}
\usepackage[]{harvard}
\usepackage{graphicx}
\usepackage{longtable}
\usepackage{rotating}
\newcommand*{\wn}{cm$^{-1}$}
\newcommand{\Dstate}{$D^{1}\Pi_{u}$}

\begin{document}

\title[The \Dstate\ state of HD]{The \Dstate\ state of HD and the mass scaling relation of its predissociation widths}
\author{G.D. Dickenson$^{1}$, W. Ubachs$^{1}$ \footnotemark[1]}
\address{$^{1}$Institute for Lasers, Life and Biophotonics Amsterdam, VU University, de Boelelaan 1081, 1081HV, Amsterdam, The Netherlands}

\begin{abstract}
Absorption spectra of HD have been recorded in the wavelength range of 75 to 90 nm at 100 K using the vacuum ultraviolet Fourier transform spectrometer at the Synchrotron SOLEIL. The present wavelength resolution represents an order of magnitude improvement over that of previous studies. We present a detailed study of the \Dstate\ - X$^{1}\Sigma^{+}_{g}$ system observed up to $v'=18$. The Q-branch transitions probing levels of $\Pi^{-}$ symmetry are observed as narrow resonances limited by the Doppler width at 100 K. Line positions for these transitions are determined to an estimated absolute accuracy of 0.06 \wn . Predissociation line widths of $\Pi^{+}$ levels are extracted from the absorption spectra. A comparison with the recent results on a study of the \Dstate\ state in H$_{2}$ and D$_{2}$ reveals that the predissociation widths scale as $\mu^{-2}J(J+1)$, with $\mu$ the reduced mass of the molecule and $J$ the rotational angular momentum quantum number, as expected from an interaction with the $B'^{1}\Sigma_{u}^{+}$ continuum causing the predissociation.
\end{abstract}

\footnotetext[1]{Corresponding Author: w.m.g.ubachs@vu.nl}

\maketitle

\section{Introduction}

Since the early years of quantum mechanics the hydrogen molecule has been studied and has provided theorist with an ideal testing ground for calculations. The stable isotopic variants of H$_{2}$, namely D$_{2}$ and HD, allow for testing mass scaling effects. The \Dstate\ state was found to undergo predissociation above the third vibrational level (the fourth vibrational level for D$_2$ and HD) which can be accurately described by Fano's theory of a single bound state interacting with a continuum~\cite{Fano1961}. 

The \Dstate\ state of HD has received considerably less interest compared to the other two stable isotopic variants. On the experimental side \citeasnoun{Takezawa1972} determined Q(1) transitions for the lowest three vibrations accurate to within a few \wn . \citeasnoun{Monfils1965} measured level energies for both $\Pi^{+}$ and $\Pi^{-}$ parity components up to $v'=8$ with accuracies of $\sim$5 \wn . A profile analysis of the predissociated line shapes was conducted by Dehmer and Chupka for $v'=7$ and 9 as well as a separate study focussing on the line positions for the R(0), R(1) and Q(1) transitions from $v'=7-16$ \cite{Dehmer1980,Dehmer1983} with accuracies of $\sim$4 \wn . Theoretically,~\citename{Kolos1976} have calculated the vibrational levels up to the $n=3$ dissociation limit for the \Dstate\ state \cite{Kolos1976} while~\citename{Abgrall2006} calculated term values for the lowest three vibrations ($v'=0-2$) in a study focussing mainly on the Lyman and Werner bands of HD~\cite{Abgrall2006}.

The present work on HD is an extension to the studies of the \Dstate\ state in H$_{2}$ and D$_{2}$ \cite{Dickenson2010a,Dickenson2011}. The measurements were obtained with the  vacuum ultraviolet (VUV) Fourier transform spectrometer (FTS) at the DESIRS beamline of the synchrotron SOLEIL. The line widths of transitions probing $\Pi^{+}$ levels for all three hydrogen isotopomers are used to verify scaling laws for the predissociation in the \Dstate\ state. 
%
%
%
\section{Experiment}
\label{sec:Experiment}

\begin{figure}[t]
\begin{center}
\includegraphics[width=1\linewidth]{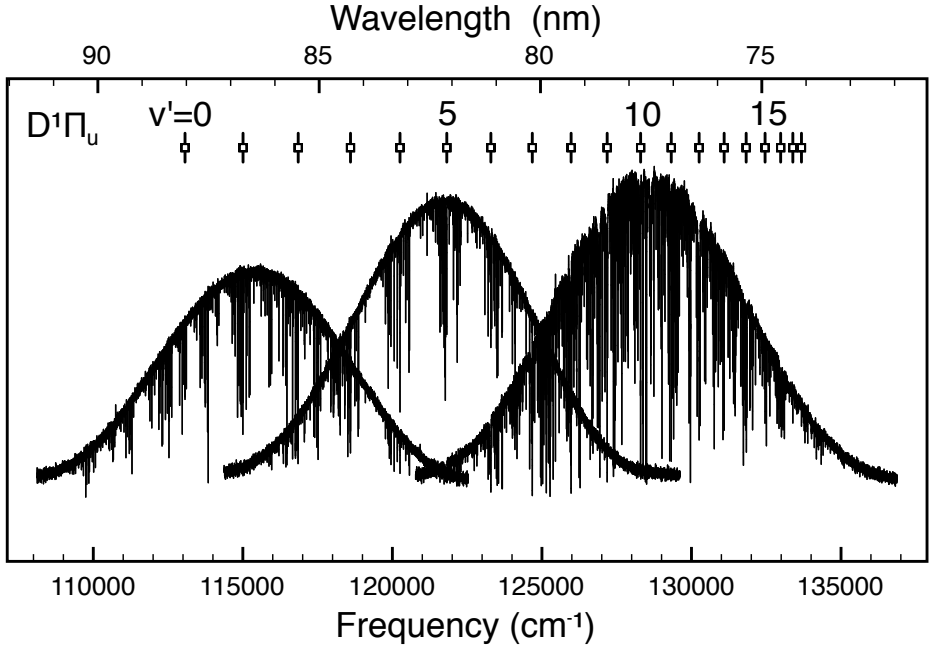}
\caption{An overview of the recorded spectra analysed in the present study. The band heads of the \Dstate\ state are indicated up to $v'=18$. Other prominent spectral features are associated with the $B^{1}\Sigma^{+}_{u}$, $B'^{1}\Sigma^{+}_{u}$ and $C^{1}\Pi_{u}$ states below the $n=2$ dissociation limit~\cite{Ivanov2010} and $B''^{1}\Sigma^{+}_{u}$ states above $n=2$. Above the ionization limit, 124568 \wn \cite{Sprecher2010} many auto-ionization resonances appear in the spectrum.}
\label{Fig:Overview}
\end{center}
\end{figure}

The VUV FTS is a scanning wavefront division interferometer operational from 40-200 nm. It has been used previously in a study of the Lyman and Werner bands of HD~\cite{Ivanov2010}. We provide only a short description of the experimental configuration, for a detailed explanation we refer to the works of~\citename{deOliveira2009} \cite{deOliveira2009,deOliveira2011}. The light source is undulator based and can be tuned continuously to produce a bell - shaped output window spanning approximately 5 nm as illustrated in figure~\ref{Fig:Overview}. The undulator radiation passes through a windowless T - shaped cell, 10 cm in length, which contains a quasi-static HD sample, slowly flowing out either side of the cell. The purity of the HD gas is estimated at $\geq$ 99\% with some traces of H$_{2}$ resulting in weak spectral features associated with H$_{2}$ Lyman bands in the low energy region. The HD is cooled to a temperature of 100 K by liquid nitrogen which flows over the outside of the T - shaped cell.

Each measurement was recorded by taking 512 kilo samples of data over the optical path difference, resulting in an instrumental width of 0.33 \wn . The final spectral windows were averaged over 100 individual interferograms and took about two hours to accumulate. The pressure inside the absorption cell can be regulated resulting in a change in the column density. Spectra were recorded at sufficient column density so that transitions appear with optimal signal-to-noise ratio but not saturated. A spectral range spanning from 112 000  - 134 000 \wn\ (75-90 nm) was covered by three spectral windows each set at a different central wavelength as shown in figure~\ref{Fig:Overview}.

The wavelength scale in the FT spectra display a strict linearity so that only one fixed point is required for a calibration. This is provided by a transition in atomic argon present in the gas filter which is used to remove higher order harmonics of the selected wavelength produced by the undulator. The transition is the $(3p)^{5}(^{2}\rm{P}_{3/2})9d([3/2]) - (3p)^{6}$ $^{1}\rm{S}_{0}$ at 125718.13 \wn\ known to an accuracy of 0.03 \wn\ \cite{Sommavilla2002}.

\section{Theory}
\label{sec:Theory}

The predissociation of the \Dstate\ state is due to a strong Coriolis coupling to the continuum of the $B'^{1}\Sigma_{u}^{+}$ state~\cite{Monfils1961}. Due to the $\Sigma^{+}$ character of the continuum, transitions probing levels of $\Pi^{-}$ symmetry are not affected by this interaction and are only very weakly predissociated due to coupling with the lower lying $C^{1}\Pi_{u}$ continuum~\cite{Glass-Maujean2010a}. A single continuum interacting with a bound state is described by Fano's theory~\cite{Fano1961} and produces broadened asymmetric absorption profiles described by the Fano function. For more details we refer to our previous works~\cite{Dickenson2010a,Dickenson2011}. The widths, broadened by life-time shortening due to the predissociation are given by
\begin{center} 
\begin{eqnarray}
\Gamma_{v'} = 2 \pi | \langle \psi_{B'\epsilon}|H(R)|\psi_{Dv'} \rangle |^{2}
\label{eqn:Widths}
\end{eqnarray}
\end{center}
where $\psi_{B'\epsilon}$ and $\psi_{D}$ are the wavefunctions of the $B'$ continuum and the discrete $D$ state respectively. Here the energy value $B'_{\epsilon}$ of the $B'$ state is taken equal to the non-perturbed energy of the discrete level $D_{v'}$. The rotational operator, $H(R)$, is the $\overrightarrow{L} \cdot \overrightarrow{R}$ operator (also responsible for the $\Lambda$-doublet splitting) which causes the predissociation widths of levels $D_{v'}$ to scale as 
\begin{center} 
\begin{eqnarray}
 \Gamma_{v'} \propto \frac{1}{\mu^{2}}J(J+1)
\label{eqn:V}
\end{eqnarray}
\end{center}
where $\mu$ is the reduced mass of the molecule and $J$ is the rotational quantum number. The reduced mass for the three isotopomers H$_{2}$, HD and D$_{2}$ are 0.5, 0.67 and 1.0 a.m.u. respectively.

\section{Results and Discussion}
\label{sec:ResultsAndDiscussion}

\begin{figure}[t]
\includegraphics[width=1\linewidth]{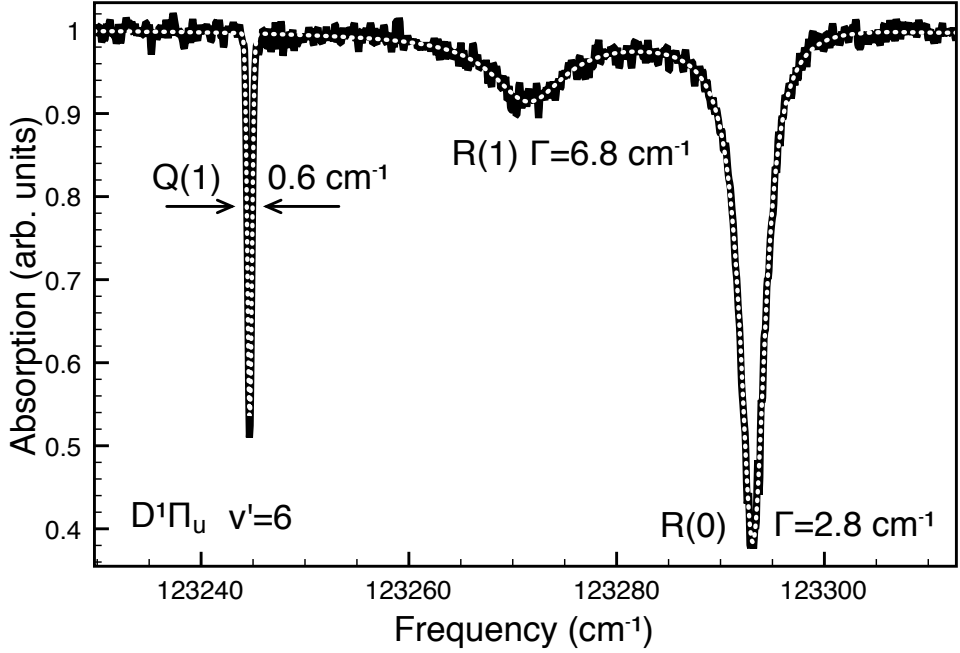}
\caption{Detailed spectrum of the $D^{1}\Pi_{u}(v'=6)$ - $X^{1}\Sigma^{+}_{g}(v''=0)$ band with R(0), R(1) and Q(1) transitions. The dotted white line represents a least squares fit of the data with the appropriate convoluted functions (see text for details).}
\label{Fig:Dv=6}
\end{figure}

The region above the second dissociation limit in HD is a complex multi-line spectrum that when cooled to liquid nitrogen temperatures consists of six overlapping Rydberg series~\cite{Dehmer1983}. The absorption spectrum is heavily congested making a complete analysis of all spectral features a challenge. In particular the \Dstate\ state is recognisable from the broadened Beutler - Fano profiles, aiding the assignment thereof. Our assignments agree with the previous works of \citeasnoun{Monfils1965} and \citeasnoun{Dehmer1983} which are accurate to within 4-5 \wn . The largest discrepancy occurs for $D^{1}\Pi_{u}(v'=9)-X^{1}\Sigma^{+}_{g}(v''=0)$ band, differing from the present line positions by $\sim 8$ \wn\ possibly attributable to wavelength drive slippage of the monochromator, as mentioned by the authors. Beyond $v'=16$ the identifications are aided by the calculations of the band heads made by \citeasnoun{Kolos1976} accurate to within 1 - 3 \wn . From an estimate of the $\Lambda$-doublet splitting the Q(1) transitions could be identified. The R(1) transitions beyond $v'=15$ were too weak to be observed.

The Q-branch transitions, observed as narrow resonances limited by Doppler broadening, were observed up to $v'=18$ and are listed in table~\ref{Tab:Pi-}. The R-branch transitions which are broadened for $v'\geq4$, were also observed up to $v'=18$ (R(0) transitions only). Transition energies and predissociated widths for these transitions are listed in table~\ref{Tab:Pi+}. All line positions and predissociated widths listed in the tables stem from a deconvolution procedure as described in previous work on D$_{2}$~\cite{Dickenson2011}. Briefly, the absorption profiles are first convoluted with a Gaussian function representative of the Doppler profile at 100 K. In a second step the Beer-Lambert law is included, accounting for the non-linear absorption depth. Finally the resulting profiles are convoluted with the instrument function and the fit parameters are then optimized by a standard least squares fitting routine. Included in the parameters are the points to an unbounded, cubic spline fit of the background. These are optimised along with the line shape parameters resulting in a fit of the background. A sample fit of the Q(1), R(1) and R(0) transitions belonging to the $D-X$ (6,0) band is shown in figure~\ref{Fig:Dv=6}. The Q(1) transition has a width of approximately 0.6 \wn\ which stems from the contribution of the instrument width of 0.33 \wn\ and the Doppler width of 0.5 \wn\ and represents the limiting resolution for the particular configuration of the FTS used.

In this analysis of the line widths the Beutler-Fano asymmetry of the line shape, represented by the Fano $q$-parameter was included by fixing the $q$ values to their theoretical prediction. Upon mass-scaling the $q$-parameters ($q \propto \mu$) it follows that $q \sim -25$ for R(0) and $q \sim -15$ for R(1) transitions in HD \cite{Glass-Maujean1979,Dickenson2011}. The present data did not permit to perform a reliable two parameter fit to extract both $q$ and $\Gamma$. This is in part due to the method of recording the spectra in absorption against a fluctuating continuum level. For further discussion see the previous work on D$_{2}$~\cite{Dickenson2011}.
 
\subsection{Spectroscopic Results}
\label{subsec:SpectroscopicResults}

\begin{figure}
\begin{center}
\includegraphics[width=1\linewidth]{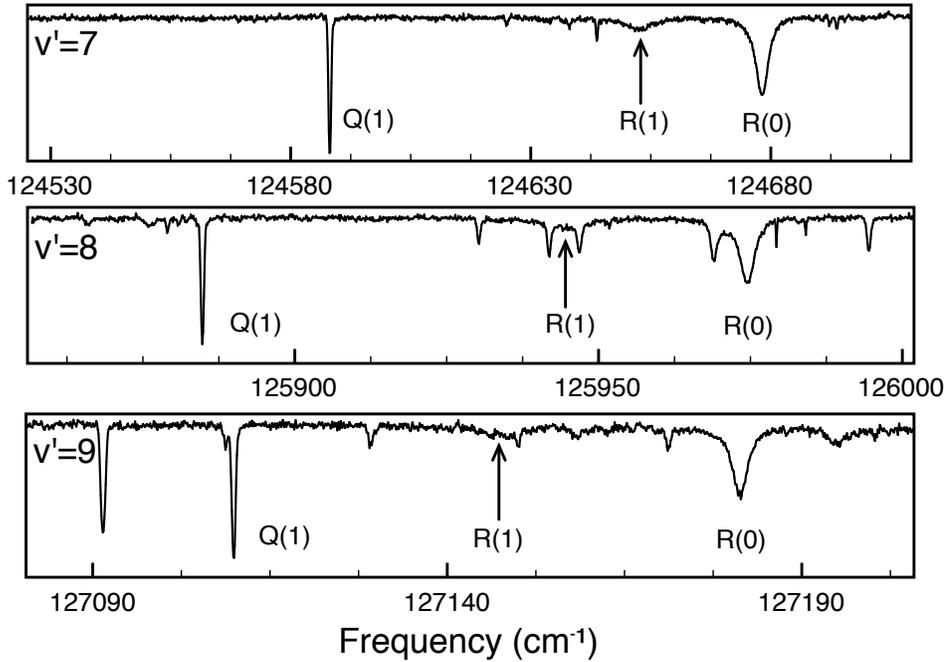}
\caption{Small portions of spectral windows containing the $D^{1}\Pi_{u}(v'=7-9)$ - $X^{1}\Sigma^{+}_{g}(v''=0)$ bands of HD. These bands are predissociated and display typical broadened Beutler-Fano profiles.}
\label{Fig:BeutlerFano}
\end{center}
\end{figure}

The Q-branch transitions, \emph{i.e.} transitions probing states of $\Pi^{-}$ symmetry, and transitions belonging to bands with $v' \leq 3$ are not predissociated and observed as narrow features with width, $\sim$0.6 \wn , equal to the Doppler width of HD at 100 K convoluted with the instrument width. Uncertainty in the reported line positions in table~\ref{Tab:Pi-} and for the unpredissociated bands listed in table~\ref{Tab:Pi+} is estimated at 0.06 \wn . For slightly saturated lines, blended lines and weak lines the uncertainty estimate increases to 0.08 \wn .

The R- and P-branch transitions are observed as broadened due to the life-time shortening caused by predissociation. Several small portions of the spectra are displayed in figure~\ref{Fig:BeutlerFano} including the $D^{1}\Pi_{u}(v'=7-9)$-$X^{1}\Sigma_{g}^{+}$ bands which display typical predissociation broadening. These transitions were fitted with convoluted profiles and the resultant line positions and widths are listed in table~\ref{Tab:Pi+}. The P(2) and R(2) transitions for bands with $v' \geq 3$ were observed as extremely weak as a result of the fact that most of the rotational population resides in the $J''=0$ and 1 levels. We estimate an uncertainty of 0.20 \wn\ on the line positions of the R(0) transitions which were observed to be $\sim$3 \wn\ broad and a 0.4 \wn\ uncertainty estimate on the R(1) transitions observed at widths of $\sim$7 \wn . There were a number of blended lines, most of which could still be fitted. Those lines affected severely by blending are indicated in the table and the estimated line position uncertainty is doubled.

\begin{figure}[t]
\begin{center}
\includegraphics[width=1\linewidth]{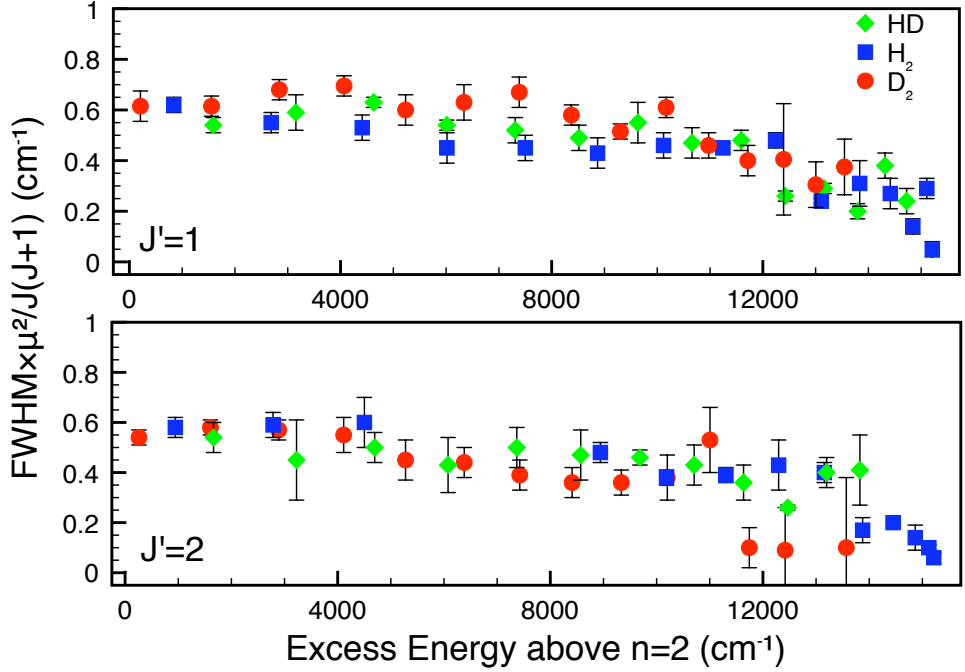}
\caption{(color online) The predissociation widths scaled by multiplying by the reduced mass squared ($\mu^{2}$). The energy scale on the $x$-axis is referenced to the $n$=2 dissociation limit of each molecule as determined by~\citeasnoun{Eyler1993}. See text for details.}
\label{Fig:Widths}
\end{center}
\end{figure}

\subsection{Predissociated Widths}
\label{subsec:PredissociatedWidths}

Figure~\ref{Fig:Widths} depicts the predissociated widths as a function of the excess energy above the $n$=2 dissociation limit for all three stable isotopomers, H$_{2}$ and D$_{2}$ as determined in previous work~\cite{Dickenson2010a,Dickenson2011} and the newly determined HD widths. The dissociation limits used for H$_{2}$, D$_{2}$ and HD were the H(1$S$)+H(2$S$), D(1$S$)+D(2$S$) and H(1$S$)+D(2$S$) respectively~\cite{Eyler1993}. The measured widths have been scaled by their respective reduced masses squared and the rotational dependence has been removed by dividing through by $J(J+1)$. The data for H$_{2}$ between 5000 and 8000 \wn\ is missing due to blending with the $B''$ state in this region. The agreement between the three isotopomers for both $J'$=1 (derived from R(0) transitions) and $J'=2$ (derived from R(1) transitions) rotational levels is good yielding further proof of the applicability of the simple two state model to the predissociation of the \Dstate\ state in all three stable hydrogen isotopomers. At the present level of accuracy the data indicates that $u-g$ symmetry breaking effects in HD do not play a role in the predissociated life-times and that the predissociation can be fully described by the $| \langle \psi_{B'\epsilon}|H(R)|\psi_{Dv'} \rangle |$ interaction.

\subsection{$\Lambda$-Doublet}
\label{subsec:LambdaDoublet}

\begin{figure}
\begin{center}
\includegraphics[width=1\linewidth]{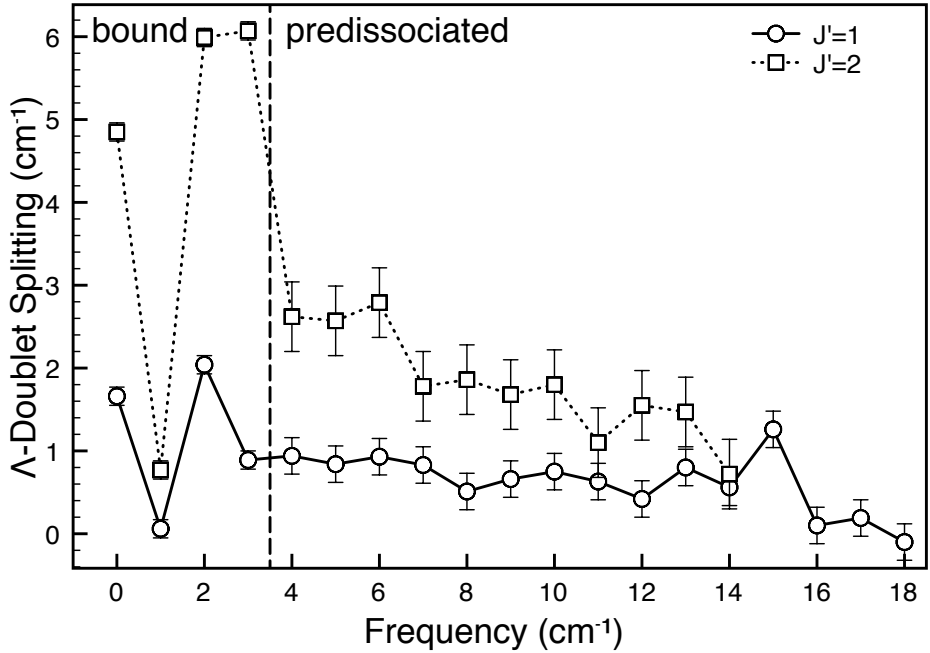}
\caption{The $\Lambda$-doublet splittings in the \Dstate\ state of HD for the $J'$=1 and 2 rotational states. The lines joining the points are to guide the eye.}
\label{Fig:LambdaDoublet}
\end{center}
\end{figure}

The $\Lambda$-doublet splitting, as depicted in figure~\ref{Fig:LambdaDoublet}, was determined by adding the ground state level energy to the Q-branch transitions~\cite{Komasa2011} and subtracting this from the R-branch transitions probing the same $J'$ but opposite $(e)-(f)$ parity. The results mirror those obtained for H$_{2}$ and D$_{2}$. The $\Lambda$-doublet splittings follow an erratic behaviour \textbf{for $v' < 4$} caused by the interactions between the discrete $B'$ and $D$ state levels. For $v'\geq4$ it follows a relatively smoothly decaying trend similar as in the observations on H$_{2}$ and D$_{2}$~\cite{Dickenson2010a,Dickenson2011}. If the assumption can be made that the $B'$ state is the sole perturber causing the $\Lambda$-doubling, the $\Lambda$-doublet splitting can be represented as,
\begin{eqnarray}
\Lambda_{v'}(J') \propto \sum_{B'v,\epsilon} \frac{\vert \langle \psi_{B'v,\epsilon} \vert H(R) \vert \psi_{Dv'} \rangle \vert^{2}}{E_{\Pi^{+}_{v}}-E_{B'v,\epsilon}} 
\end{eqnarray}
where summation over all $B'$ levels includes the bound levels below $n=2$ and an integral over the $B'$ continuum. Interaction with the $B'$ levels of $^{1}\Sigma^{+}$ symmetry causes the $\Pi^{+}(e)$ levels of the \Dstate\ state to shift upward while the $\Pi^{-}(f)$ levels are unaffected. Similarly as in the deviation of the predissociation widths the $\Lambda$-doublet splitting then scales like $\mu^{-2}J(J+1)$. The present results on HD and the results of H$_{2}$~\cite{Dickenson2010a} perfectly match this scaling, while the $\Lambda$-doublet splittings in D$_{2}$~\cite{Dickenson2011} are somewhat too large in this comparison.
\section{Conclusion}

The VUV FTS observations on the \Dstate\ state have been extended to HD. The present work represents the highest resolution study on this state performed so far. The predissociated line shapes were analysed resulting in predissociated line widths determined to a high level of accuracy. The present and previous studies show through the mass scaling and rotational scaling that the predissociation in the $\Pi^{+}$ parity states of the \Dstate\ state can be modelled by a rotational interaction to the continuum of the $B'^{1}\Sigma^{+}_{u}$ state. In the case of HD the $u-g$ symmetry breaking does not play a role in the predissociated widths at the present level of accuracy.
\section{Acknowledgements}

GDD is grateful to the SOLEIL staff scientists L. Nahon, N. de Oliveira and D. Joyeux for the hospitality and for the collaboration. Dr A. Heays is thanked for valuable advice regarding the fit model. The EU provided financial support through the transnational funding scheme. This work was supported by the Netherlands Foundation for Fundamental Research of Matter (FOM). The authors thank two anonymous referees for valuable suggestions.

\section*{References}

\bibliographystyle{jphysicsB}
\bibliography{/Users/garydee/Documents/Articles/CompleteDataBase}

\clearpage

\begin{longtable}{l r @{.} l r @{.} l l r @{.} l r @{.} l}
\caption[]{Transition frequencies of Q - branch transitions probing levels of $\Pi^{-}$ symmetry. $\Delta$ represents a comparison with the previous measurements of \citeasnoun{Monfils1965} for $v'=0-6$ and \citeasnoun{Dehmer1983} for $v'=7-16$, defined as the present measurement minus the previous measurements. All values in \wn .}
\label{Tab:Pi-}\\
\hline
\multicolumn{3}{c}{Transition Frequency} & \multicolumn{2}{c}{$\Delta$} &\multicolumn{3}{c}{Transition Frequency} & \multicolumn{2}{c}{$\Delta$}     \\
\hline
\\
\endfirsthead
\hline
\multicolumn{3}{c}{Transition Frequency} & \multicolumn{2}{c}{$\Delta$}&\multicolumn{3}{c}{Transition Frequency} & \multicolumn{2}{c}{$\Delta$}   \\ 
\hline
\\
\endhead

\hline
\multicolumn{8}{c}{{continued on next page}} \\ 
\endfoot

\hline
\endlastfoot

	\multicolumn{5}{c}{$D-X(0,0)$}						&				\multicolumn{5}{c}{$D-X(1,0)$}								\\
Q(1)$^{s}$	&	112975	&	18	&	1	&	11	&	Q(1)$^{s}$	&	114916	&	47	&	5	&	10	\\
Q(2)	&	112886	&	03	&	1	&	35	&	Q(2)	&	114822	&	09	&	-0	&	07	\\
Q(3)	&	112753	&	22	&	1	&	77	&	Q(3)	&	114684	&	45	&	2	&	26	\\
	\multicolumn{5}{c}{$D-X(2,0)$}						&				\multicolumn{5}{c}{$D-X(3,0)$}								\\
Q(1)$^{s}$	&	116760	&	16	&	2	&	73	&	Q(1)$^{s}$	&	118508	&	76	&	2	&	42	\\
Q(2)$^{s}$	&	116663	&	06	&	1	&	69	&	Q(2)$^{s}$	&	118407	&	81	&	2	&	58	\\
Q(3)	&	116518	&	29	&	1	&	36	&	Q(3)	&	118256	&	35	&	1	&	39	\\
	\multicolumn{5}{c}{$D-X(4,0)$}						&				\multicolumn{5}{c}{$D-X(5,0)$}								\\
Q(1)$^{s}$	&	120164	&	47	&	-1	&	01	&	Q(1)	&	121729	&	21	&	0	&	90	\\
Q(2)	&	120059	&	79	&	0	&	96	&	Q(2)	&	121620	&	99	&	0	&	99	\\
Q(3)	&	119900	&	37	&	0	&	46	&	Q(3)	&	121459	&	67	&	1	&	20	\\
	\multicolumn{5}{c}{$D-X(6,0)$}						&				\multicolumn{5}{c}{$D-X(7,0)$}								\\
Q(1)	&	123203	&	12	&	1	&	95	&	Q(1)	&	124588	&	35	&	1	&	25	\\
Q(2)	&	123091	&	11	&	1	&	09	&	Q(2)	&	124472	&	72	\\
Q(3)	&	122924	&	14	&	0	&	69	&	Q(3)	&	124300	&	33	\\
	\multicolumn{5}{c}{$D-X(8,0)$}						&				\multicolumn{5}{c}{$D-X(9,0)$}								\\
Q(1)	&	125885	&	02	&	1	&	62	&	Q(1)	&	127091	&	67	&	-7	&	23	\\
Q(2)	&	125765	&	66	&	\multicolumn{2}{c}{} &	Q(2)	&	126968	&	71	\\
	\multicolumn{5}{c}{$D-X(10,0)$}						&				\multicolumn{5}{c}{$D-X(11,0)$}								\\
Q(1)	&	128208	&	79	&	5	&	29	&	Q(1)	&	129234	&	35	&	1	&	15	\\
Q(2)	&	128082	&	25	&\multicolumn{2}{c}{}&	Q(2)	&	129103	&	84	\\
	\multicolumn{5}{c}{$D-X(12,0)$}						&				\multicolumn{5}{c}{$D-X(13,0)$}								\\
Q(1)	&	130166	&	57	&	-0	&	43	&	Q(1)	&	131002	&	36	&	2	&	56	\\
Q(2)	&	130032	&	21	&\multicolumn{2}{c}{}&	Q(2)	&	130864	&	00		\\
	\multicolumn{5}{c}{$D-X(14,0)$}						&				\multicolumn{5}{c}{$D-X(15,0)$}								\\
Q(1)$^{b}$	&	131737	&	82	&	-1	&	88	&	Q(1)	&	132369	&	36	&	3	&	66	\\
Q(2)	&	131595	&	19	\\
	\multicolumn{5}{c}{$D-X(16,0)$}						&				\multicolumn{5}{c}{$D-X(17,0)$}								\\
Q(1)$^{b}$	&	132891	&	54	&	-0	&	06	&	Q(1)	&	133299	&	89\\
	\multicolumn{5}{c}{$D-X(18,0)$}																		\\
Q(1)	&	133588	&	59															\\
																			\\

\\
\\
\footnotesize
$^{s}$\emph{Saturated}\\
\footnotesize
$^{b}$\emph{Blended}\\
\\
\end{longtable}

\begin{longtable}{l r @{.} l l r @{.} l l r @{.} l l r @{.} l}
\caption[]{Transition frequencies of R - branch transitions probing levels of $\Pi^{+}$ symmetry. $\Delta$ represents a comparison with the previous measurements of \citeasnoun{Monfils1965} for $v'=0-6$ and \citeasnoun{Dehmer1983} for $v'=7-16$ defined as the present measurement minus the previous measurements. $\Gamma$ represents the predissociation width. All values in \wn .}
\label{Tab:Pi+}\\
\hline
\multicolumn{3}{l}{Transition Frequency} & \multicolumn{1}{c}{$\Gamma$}& \multicolumn{2}{c}{$\Delta$} &\multicolumn{3}{l}{Transition Frequency} & \multicolumn{1}{c}{$\Gamma$}&\multicolumn{2}{c}{$\Delta$}     \\
\hline
\\
\endfirsthead
\hline
\multicolumn{3}{l}{Transition Frequency} & \multicolumn{1}{c}{$\Gamma$}&\multicolumn{2}{c}{$\Delta$} &\multicolumn{3}{l}{Transition Frequency} & \multicolumn{1}{c}{$\Gamma$}&\multicolumn{2}{c}{$\Delta$}   \\ 
\hline
\\
\endhead

\hline
\multicolumn{8}{c}{{continued on next page}} \\ 
\endfoot

\hline
\endlastfoot
	\multicolumn{6}{c}{$D-X(0,0)$}						&						\multicolumn{6}{c}{$D-X(1,0)$}										\\
 R(0)$^{s}$	&	113066	&	07	&		&	-0	&	75	&	 R(0)$^{s}$	&	115005	&	76	&		&	1	&	17	\\
 R(1)$^{s}$	&	113068	&	72	&		&	2	&	84	&	 R(1)$^{s}$	&	115000	&	70	&		&	-4	&	21	\\
	\multicolumn{6}{c}{$D-X(2,0)$}						&						\multicolumn{6}{c}{$D-X(3,0)$}										\\
 R(0)$^{s}$	&	116851	&	42	&		&	4	&	44	&	 R(0)$^{s}$	&	118598	&	93	&		&	0	&	79	\\
 R(1)$^{s}$	&	116846	&	89	&		&	1	&	53	&	 R(1)$^{s}$	&	118591	&	70	&		&	2	&	86	\\
	\multicolumn{6}{c}{$D-X(4,0)$}						&						\multicolumn{6}{c}{$D-X(5,0)$}										\\
 R(0)	&	120254	&	63	&	2.4	&	0	&	08	&	 R(0)	&	121819	&	48	&	2.6	&	0	&	78	\\
 R(1)	&	120240	&	26	&	7.3	&	-1	&	05	&	 R(1)	&	121801	&	44	&	6.0	&	2	&	62	\\
	\multicolumn{6}{c}{$D-X(6,0)$}						&						\multicolumn{6}{c}{$D-X(7,0)$}										\\
 R(0)	&	123293	&	27	&	2.8	&	0	&	43	&	 R(0)	&	124678	&	40	&	2.4	&	0	&	70	\\
 R(1)	&	123271	&	74	&	6.8	&	0	&	91	&	 R(1)	&	124652	&	34	&	5.8	&	0	&	56	\\
	\multicolumn{6}{c}{$D-X(8,0)$}						&						\multicolumn{6}{c}{$D-X(9,0)$}										\\
 R(0)	&	125974	&	74	&	2.3	&	1	&	15	&	 R(0)	&	127181	&	58	&	2.2	&	-7	&	82	\\
 R(1)	&	125945	&	60	&	6.8	&	0	&	09	&	 R(1)	&	127148	&	23	&	6.4	&	-8	&	97	\\
	\multicolumn{6}{c}{$D-X(10,0)$}						&						\multicolumn{6}{c}{$D-X(11,0)$}										\\
 R(0)	&	128298	&	77	&	2.5	&	3	&	67	&	 R(0)	&	129324	&	08	&	2.1	&	2	&	28	\\
 R(1)	&	128261	&	88	&	6.2	&	3	&	18	&	 R(1)	&	129282	&	73	&	5.9	&	1	&	23	\\
	\multicolumn{6}{c}{$D-X(12,0)$}						&						\multicolumn{6}{c}{$D-X(13,0)$}										\\
 R(0)$^{b}$	&	130256	&	22	&	2.1	&	-3	&	38	&	 R(0)	&	131092	&	40	&	1.2	&	1	&	60	\\
 R(1)	&	130211	&	60	&	4.8	&	-1	&	70	&	 R(1)$^{b}$	&	131043	&	31	&	3.5	&	2	&	31	\\
	\multicolumn{6}{c}{$D-X(14,0)$}						&						\multicolumn{6}{c}{$D-X(15,0)$}										\\
 R(0)	&	131827	&	56	&	1.3	&	-1	&	54	&	 R(0)$^{b}$	&	132459	&	67	&	0.9	&	3	&	37	\\
 R(1)	&	131773	&	73	&	5.4	&	-2	&	57	&	 R(1)	&	132400	&	58	&	5.5	\\
	\multicolumn{6}{c}{$D-X(16,0)$}						&						\multicolumn{6}{c}{$D-X(17,0)$}										\\
 R(0)	&	132980	&	87	&	1.7	&	-1	&	53	&	 R(0)	&	133389	&	31	&	1.1\\
	\multicolumn{6}{c}{$D-X(18,0)$}																						\\
 R(0)$^{b}$	&	133677	&	72																\\
\\
\\
\footnotesize
$^{s}$\emph{Saturated}\\
\footnotesize
$^{b}$\emph{Blended}\\
\\
\end{longtable}

\end{document}